\documentclass [10pt]{article}
\usepackage[left=2cm,top=2.50cm,right=2cm,bottom=2.50cm]{geometry}
\usepackage{mathrsfs}
\usepackage{amsmath,amssymb,latexsym,color,cancel,graphicx,bbm,colortbl}
\usepackage[english]{babel}
\usepackage[latin1]{inputenc}
\usepackage{ragged2e}
\usepackage{cite}

\begin{document}
\date{}

\title{Berry phase of the Tavis-Cummings model with three modes of oscillation}
\author{E. Chore\~no$^{a}$, D. Ojeda-Guill\'en$^{a,b}$\footnote{{\it E-mail address:} dojedag@ipn.mx}, R. Valencia$^{a}$\\ and V. D. Granados$^{a}$}
\maketitle

\begin{minipage}{0.9\textwidth}
\small $^{a}$ Escuela Superior de F{\'i}sica y Matem\'aticas,
Instituto Polit\'ecnico Nacional, Ed. 9, Unidad Profesional Adolfo L\'opez Mateos, Delegaci\'on Gustavo A. Madero, C.P. 07738, Ciudad de M\'exico, Mexico.\\
\small $^{b}$ Escuela Superior de C\'omputo, Instituto Polit\'ecnico Nacional,
Av. Juan de Dios B\'atiz esq. Av. Miguel Oth\'on de Mendiz\'abal, Col. Lindavista,
Delegaci\'on Gustavo A. Madero, C.P. 07738, Ciudad de M\'exico, Mexico.\\

\end{minipage}

\begin{abstract}
In this paper we develop a general method to obtain the Berry phase of time-dependent Hamiltonians with a linear structure
given in terms of the $SU(1,1)$ and $SU(2)$ groups. This method is based on the similarity transformations of the displacement
operator performed to the generators of each group, and let us diagonalize these Hamiltonians. Then, we introduce a trilinear
form of the Tavis-Cummings model to compute the $SU(1,1)$ and $SU(2)$ Berry phases of this model.
\end{abstract}

\section{Introduction}

The Jaynes-Cummings model is the simplest and completely soluble quantum-mechanical model which describes the interaction between
radiation and matter \cite{Jay}. The exact solution of this theoretical model has been found in the rotating wave approximation \cite{Haroche}.
However, despite the simplicity of the Jaynes-Cummings model, it presents interesting quantum phenomena \cite{Narozhny,Kuklinski,Short,Diedrich,Milonni,Slosser,Gea,Phoenix}, all of them being experimentally corroborated \cite{Goy,Brune,Guerlin}.

The Tavis-Cummings model is another important model in quantum optics, which emerged from the study of $N$ identical two-level molecules interacting through a dipole coupling with a single-mode quantized radiation field at resonance \cite{Dicke,TC}. This model has been studied in terms of the Holstein-Primakoff transformation \cite{Bashir}, quantum inverse methods \cite{Bogoliubov,Rybin}, and polynomially deformed $su(2)$ algebras \cite{Vadeiko}.
In general, the Jaynes-Cummings model and the Tavis-Cummings model are still under study as can be seen in the references \cite{Lamata,Gerritsma,Lamata2,Retzker,Kopylov,Sun,Nos1,Nos2,Nos3,Nos4}.

On the other hand, since its introduction in 1984 the Berry phase \cite{Berry} has been extensively studied in several quantum systems \cite{Gerry,Xie,Bu,Thilagam}. This is a phase factor gained by the wavefunction after the system is transported through a closed path via adiabatic variation of parameters. The aim of the present work is to compute the Berry phase of the Tavis-Cummings model with three modes of oscillation in terms of the $SU(1,1)$ and $SU(2)$ group theory.

This work is organized as follows. In Section 2, we introduce a general method to diagonalize a Hamiltonian with a liner $su(1,1)$ and $su(2)$ linear structure by means of the tilting transformation of each group. In Section 3, we compute the Berry phase for time-dependent Hamiltonians with a linear structure given in terms of the $SU(1,1)$ and $SU(2)$ groups. With all these previous results, Section 4 is dedicated to obtain the $SU(1,1)$ and $SU(2)$ Berry phases of the Tavis-Cummings model with three modes of oscillation.

\section{Algebraic diagonalization method of a system with an $SU(1,1)$ or $SU(2)$ symmetry}

As it is well known, the group algebraic methods are very powerful tools in the description, diagonalization as well as understanding of the nature of the physical structure of many problems with certain dynamical symmetry. In this Section, we outline an algebraic procedure for diagonalizing certain Hamiltonians of  physical systems that can be described through a linear structure of the Lie algebra, that is, Hamiltonians that can be written as a linear combination of the generators $\{N_{i},A_{q},A^{\dag}_{q}\}$ of the $G$ Lie algebra, i.e.
\begin{equation}
H=\sum_{i}a_{i}N_{i}+\sum_{q}\left(b_{q}A_{q}+c_{q}A^{\dag}_{q}\right),\label{GH}
\end{equation}
where $a_{i}$, $b_{q}$ and $c_{q}$ can be real or complex constants. The generators $\{N_{i},A_{q},A^{\dag}_{q}\}$ form the standard Cartan-Weyl basis of semi-simple $G$ Lie algebra satisfying the commutation relations $\left[N_{i},N_{j}\right]=0$, $\left[N_{i},A_{q}\right]=q_{i}A_{q}$, $\left[A_{q},A^{\dag}_{q}\right]=q^{i}N_{i}$, $\left[A_{p},A_{q}\right]=C_{p,q}A_{p+q}$ $\left( p\neq q\right)$. Some examples of this algebra can be constructed by combining the bosonic realizations of the generators $su(1,1)$ and $su(2)$ Lie algebras or by the bosonic generators of the $Sp(4,R)$ algebra.

In general, based on a $G$ Lie algebra with its respective generators $\{N_{i},A_{q},A^{\dag}_{q}\}$ which satisfy certain commutation relations, we can introduce a unitary operator
\begin{equation}
D(\xi_{q})=\exp\left[\sum_{q}\left(\xi_{q}A_{q}-\xi^{*}_{q}A^{\dag}_{q}\right)\right].
\end{equation}
Here, $D(\xi_{q})$ is a generalized displacement operator and $\xi_{q}$ are complex parameters to be determined. Now, by using the Baker-Campbell-Hausdorff identity
\begin{equation*}
e^{-A}Be^{A}=B+[B,A]+\frac{1}{2!}[[B,A],A]+\frac{1}{3!}[[[B,A]A]A]+...,\label{BCH}
\end{equation*}
and a suitable choice of the complex parameters $\xi_{q}$, the operator $D(\xi_{q})$ allows us to compute
the transformation of the Lie algebra generators $\{N_{i},A_{q},A^{\dag}_{q}\}$ as a linear combination of them. Hence, a generator $A_{j}$ of the Lie algebra can be transformed as
\begin{equation}
D^{\dag}(\xi_{q})A_{j}D(\xi_{q})=\sum_{i}\lambda_{ij}(\xi_{q})A_{i},
\end{equation}
where each coefficient $\lambda_{ij}$ is required to be an analytical function of the complex parameters $\xi_{q}$. From this, we have that a Hamiltonian with a semi-simple Lie algebra structure as in equation (\ref{GH}) can be transformed by $D(\xi_{q})$ to
\begin{equation}
D^{\dag}(\xi_{q})HD(\xi_{q})=\sum_{i}\Lambda_{i}(\xi,a,b,c)N_{i}+\sum_{q}\left(F_{q}(\xi,a,b,c)A_{q}+M_{q}(\xi,a,b,c)A^{\dag}_{q}\right).\label{TGH}
\end{equation}
Since the operators $N_{i}$ commute, the Hamiltonian of equation (\ref{GH}) can be diagonalized by this transformation if we impose
\begin{equation}
F_{q}(\xi,a,b,c)=0,\quad\quad M_{q}(\xi,a,b,c)=0.\label{Eq}
\end{equation}
The above expressions are a set of equations of the parameters $\xi_{i}$ and the physical constants $a_{i}$, $b_{q}$ and $c_{q}$ of the Hamiltonian. The solution of the system of equations reduces the expression (\ref{TGH}) to
\begin{equation}
D^{\dag}(\xi)HD(\xi)=\sum_{i}\Lambda_{i}(\xi,a,b,c)N_{i}.\label{Di}
\end{equation}

Therefore, if the eigenfunctions $\Phi_{n}$ of the operator $N_{i}$ ($N_{i}\Phi_{n}=\alpha_{i}\Phi_{n}$) are known we have from equation (\ref{Di}) that
\begin{align}
HD(\xi_{q})\Phi_{n}&=\sum_{i}\Lambda_{i}(\xi,a,b,c)\alpha_{i}D(\xi_{q})\Phi_{n},\\ &
H\Phi'^{(q)}_{n}=\Omega(\xi,a,b,c,\alpha_{i})\Phi'^{(q)}_{n},
\end{align}
where $\Phi'^{(q)}_{n}=D(\xi_{q})\Phi_{n}$ and $\Omega(\xi,a,b,c,\alpha_{i})=\sum_{i}\Lambda_{i}(\xi,a,b,c)\alpha_{i}$ are respectively the eigenfunctions and eigenvalues of the Hamiltonian of equation (\ref{GH}). It is important to point out that the diagonalization of the Hamiltonian $H$ depends on whether the system of equations (\ref{Eq}) have a solution.

\subsection{Diagonalization of Hamiltonians with a linear $su(1,1)$ and $su(2)$ algebraic structure}

During the last decade much attention has been given to studying different quantum optical models. As a result, it has been found that a wide variety of these models are expressed in terms of bosons and fermions or matrix-differential equations. By choosing an appropriate realization, many of these models can be put in the context of the $su(1,1)$ and $su(2)$ Lie algebras, as it is shown in the references \cite{Koc1,Koc2,Koc3}. In this Section, as an simple but useful application of the theory developed earlier, we are going to diagonalize two Hamiltonians with a simple structure given in terms of the $su(1,1)$ and $su(2)$ Lie algebras. Thus, we introduce two Hamiltonians written as a linear combination of the generators $\{K_{\pm}$, $K_{0}\}$ of $SU(1,1)$ group and the generators $\{J_{\pm}$,$K_{0}\}$ of $SU(2)$ group as
\begin{equation}
H_{su(1,1)}=a_{0}K_{0}+a_{1}K_{+}+a_{2}K_{-},\quad\quad H_{su(2)}=b_{0}J_{0}+b_{1}J_{+}+b_{2}J_{-}.\label{H's}
\end{equation}
The generators of each group satisfy the commutation relations \cite{Vourdas}
\begin{eqnarray}
[K_{0},K_{\pm}]=\pm K_{\pm},\quad\quad [K_{-},K_{+}]=2K_{0},\label{algebra1}
\end{eqnarray}
\begin{eqnarray}
[J_{0},J_{\pm}]=\pm J_{\pm},\quad\quad [J_{+},J_{-}]=2J_{0}.\label{algebra2}
\end{eqnarray}

The displacement operators $D(\xi_{1})$ and $D(\xi_{2})$ for these algebras are defined in terms of the creation and annihilation operators $\{K_+, K_- \}$ and $\{J_+,J_- \}$ as
\begin{equation}
D(\xi_{1})_{su(1,1)}=\exp(\xi_{1}K_{+}-\xi^{*}_{1}K_{-}),\quad\quad D(\xi_{2})_{su(2)}=\exp(\xi_{2}J_{+}-\xi^{*}_{2}J_{-}),\label{do}
\end{equation}
where $\xi_{i}=-\frac{1}{2}\tau_{i}e^{-i\varphi_{i}}$, $-\infty<\tau_{i}<\infty$ and $0\leq\varphi_{i}\leq2\pi$. With these operators we can transform the generators of the $su(1,1)$ and $su(2)$ Lie algebras as $K'_{s}=D^{\dag}(\xi_{1})K_{s}D(\xi_{1})$ and $J'_{s}=D^{\dag}(\xi_{2})J_{s}D(\xi_{2})$. Explicitly we obtain
\begin{align}
&K'_{+}=\frac{\xi_{1}^{*}}{|\xi_{1}|}\alpha K_{0}+(\beta+1)K_{+}+\beta\frac{\xi_{1}^{*}}{\xi_{1}}K_{-},\quad\quad J'_{+}=-\frac{\xi^{*}_{2}}{|\xi_{2}|}\delta J_{0}+(\epsilon+1)J_{+}+\epsilon\frac{\xi^{*}_{2}}{\xi_{2}}J_{-},\\& K'_{-}=\frac{\xi_{1}}{|\xi_{1}|}\alpha K_{0}+(\beta+1)K_{-}+\beta\frac{\xi_{1}}{\xi^{*}_{1}}K_{+},\quad\quad J'_{-}=-\frac{\xi_{2}}{|\xi_{2}|}\delta J_{0}+(\epsilon+1)J_{-}+\epsilon\frac{\xi_{2}}{\xi^{*}_{2}}J_{+},\\&K'_{0}=(2\beta+1)K_{0}+\frac{\alpha\xi_{1}}{2|\xi_{1}|}K_{+}+\frac{\alpha\xi^{*}_{1}}{2|\xi_{1}|}K_{-},\quad\quad J'_{0}=(2\epsilon+1)J_{0}+\frac{\delta\xi_{2}}{2|\xi_{2}|}J_{+}+\frac{\delta\xi^{*}_{2}}{2|\xi_{2}|}J_{-}.\label{transformations}
\end{align}
where $\alpha=\sinh(2|\xi_{1}|)$,$\beta=\frac{1}{2}\left[\cosh(2|\xi_{1}|)-1\right]$, $\delta=\sin(2|\xi_{2}|)$ and $\epsilon=\frac{1}{2}\left[\cos(2|\xi_{2}|)-1\right]$.

Therefore, the Hamiltonians of equation (\ref{H's}) are transformed in terms of the displacement operators $D(\xi_{1})$ and $D(\xi_{2})$ to
\begin{equation}
H'_{su(1,1)}=A_{0}(\xi_{1})K_{0}+A_{1}(\xi_{1})K_{+}+A_{2}(\xi_{1})K_{-},\quad\quad H'_{su(2)}=B_{0}(\xi_{2})J_{0}+B_{1}(\xi_{2})J_{+}+B_{2}(\xi_{2})J_{-},\label{TH's}
\end{equation}
where new $A's$ and $B's$ constants are given by
\begin{align}
&A_{0}=(2\beta+1)a_{0}+\frac{\xi^{*}_{1}}{|\xi_{1}|}\alpha a_{1}+\frac{\xi_{1}}{|\xi_{1}|}\alpha a_{2},\quad\quad B_{0}=(2\epsilon+1)b_{0}-\frac{\xi^{*}_{2}}{|\xi_{2}|}\delta b_{1}-\frac{\xi_{2}}{|\xi_{2}|}\delta b_{2},\\& A_{1}=\frac{\xi_{1}}{2|\xi_{1}|}\alpha a_{0}+(\beta+1)a_{1}+\frac{\xi_{1}}{\xi^{*}_{1}}\beta a_{2},\quad\quad B_{1}=\frac{\xi}{2|\xi_{2}|}\delta b_{0}+(\epsilon+1)b_{1}+\frac{\xi_{2}}{\xi^{*}_{2}}\epsilon b_{2},\\& A_{2}=\frac{\xi^{*}_{1}}{2|\xi_{1}|}\alpha a_{0}+\frac{\xi^{*}_{1}}{\xi_{1}}\beta a_{1}+(\beta+1)a_{2}\quad\quad\quad B_{2}=\frac{\xi^{*}_{2}}{2|\xi_{2}|}\delta b_{0}+\frac{\xi^{*}_{2}}{\xi_{2}}\epsilon b_{1}+(\epsilon+1)b_{2}.\label{AB}
\end{align}
The generators $K_{\pm}$ and $J_{\pm}$ can be removed from the Hamiltonians $H'_ {su (1,1)}$ and $H'_{su(2)}$ if we impose that coefficients $A_{1}=0$, $A_{2}=0$, $B_{1}=0$, and $B_{2}=0$. From this condition we need to solve the following system of equations
\begin{align}
&\frac{\xi_{1}}{2|\xi_{1}|}\alpha a_{0}+(\beta+1)a_{1}+\frac{\xi_{1}}{\xi^{*}_{1}}\beta a_{2}=0,\\& \frac{\xi^{*}_{1}}{2|\xi_{1}|}\alpha a_{0}+\frac{\xi^{*}_{1}}{\xi_{1}}\beta a_{1}+(\beta+1)a_{2}=0,
\end{align}
and
\begin{align}
&\frac{\xi}{2|\xi_{2}|}\delta b_{0}+(\epsilon+1)b_{1}+\frac{\xi_{2}}{\xi^{*}_{2}}\epsilon b_{2}=0,\\&  \frac{\xi^{*}_{2}}{2|\xi_{2}|}\delta b_{0}+\frac{\xi^{*}_{2}}{\xi_{2}}\epsilon b_{1}+(\epsilon+1)b_{2}=0.
\end{align}
Therefore, by choosing the coherent parameters $\tau_{i}$ and $\varphi_{i}$ of the complex numbers $\xi_{i}=-\frac{\tau_{i}}{2}e^{-i\varphi_{i}}$ as
\begin{equation}
\tau_{1}=\tanh^{-1}\left[\frac{2\sqrt{a_{1}a_{2}}}{a_{0}}\right],\quad\quad\varphi_{1}=\frac{i}{2}\ln\left[\frac{a_{1}}{a_{2}}\right]\label{parame1} ,
\end{equation}
and
\begin{equation}
\tau_{2}=\arctan\left(\frac{2\sqrt{b_{1}b_{2}}}{b_{0}}\right),\quad\quad\varphi_{2}=\frac{i}{2}\ln\left[\frac{b_{1}}{b_{2}}\right]\label{parame2},
\end{equation}
we diagonalize the Hamiltonians of equation (\ref{TH's}) to obtain
\begin{equation}
H'_{su(1,1)}=\sqrt{a_{0}^{2}-4a_{1}a_{2}}K_{0},\quad\quad H'_{su(2)}=\sqrt{a_{0}^{2}+4a_{1}a_{2}}J_{0}\label{ADHsu11}
\end{equation}
Finally, if the eigenfunctions $\Phi^{(1)}_{n}$ and $\Phi^{(2)}_{n}$ of $K_{0}$ and $J_{0}$ are known, we have found the eigenvalues of the Hamiltonians of equation (\ref{H's}). The eigenfunctions are obtained by applying the operators  $D(\xi_{i})$ on the functions $\Phi^{(i)}_{n}$. Hence, the eigenfunctions of the Hamiltonian $H_{su (1,1)}$ are given by the $SU(1,1)$ Perelomov number coherent states \cite{Nos5}
\begin{eqnarray}
|\zeta_{1},k,n\rangle &=&\sum_{s=0}^\infty\frac{\zeta_{1}^s}{s!}\sum_{j=0}^n\frac{(-\zeta_{1}^*)^j}{j!}e^{\eta_{1}(k+n-j)}
\frac{\sqrt{\Gamma(2k+n)\Gamma(2k+n-j+s)}}{\Gamma(2k+n-j)}\nonumber\\
&&\times\frac{\sqrt{\Gamma(n+1)\Gamma(n-j+s+1)}}{\Gamma(n-j+1)}|k,n-j+s\rangle.\label{PNCS}
\end{eqnarray}
Similarly, the eigenfunctions of the Hamiltonian $H_{su (2)}$ are given by the $SU(2)$ Perelomov number coherent states \cite{Nos5}
\begin{eqnarray}
|\zeta_{2},j,\mu\rangle &=&\sum_{s=0}^{j-\mu+n}\frac{\zeta_{2}^{s}}{s!}\sum_{n=0}^{\mu+j}\frac{(-\zeta_{2}^*)^{n}}{n!}e^{\eta_{2}(\mu-n)}
\frac{\Gamma(j-\mu+n+1)}{\Gamma(j+\mu-n+1)}\nonumber\\ &&\times\left[\frac{\Gamma(j+\mu+1)\Gamma(j+\mu-n+s+1)}{\Gamma(j-\mu+1)\Gamma(j-\mu+n-s+1)}\right]^{\frac{1}{2}}|j,\mu-n+s\rangle,\label{PNCS2}
\end{eqnarray}
where $\zeta_{1}=-\tanh(\frac{\tau_{1}}{2})e^{-i\varphi_{1}}$, $\eta_{1}=\ln(1-|\zeta_{1}|^{2})$ and $\zeta_{2}=-\tan(\frac{\tau_{2}}{2})e^{-i\varphi_{2}}$ and $\eta_{2}=\ln(1+|\zeta_{2}|^{2})$.

Therefore, it is convenient that $K_{0}$ and $J_{0}$ be operators such that we know their eigenfunctions and eigenvalues. Moreover, notice that if the coefficients $a_{0}$, $a_{1}$ and $a_{2}$ are equal, the problem does not have an exact solution.

\section{Time-dependent Hamiltonians with a linear $su(1,1)$ or $su(2)$ Lie algebraic structure}

In this Section, we shall consider systems whose Hamiltonian $H(t)$ is an explicit function of time and has a linear $su(1,1)$ or $su(2)$ Lie algebra structure, i.e.,
\begin{equation}
H(t)_{su(1,1)}=a_{0}(t)K_{0}+a_{1}(t)K_{+}+a_{2}(t)K_{-},\quad\quad H(t)_{su(2)}=b_{0}(t)J_{0}+b_{1}(t)J_{+}+b_{2}(t)J_{-}.
\end{equation}
Now, because the Hamiltonians $H(t)_{su(1,1)}$ and $H(t)_{su(2)}$ are Hermitian operators, we have $a_{2}(t)=a^{*}_{1}(t)$, $b_{2}(t)=b^{*}_{1}(t)$. Moreover, we can write the coefficients $a_{1}(t)$ and $b_{1}(t)$ as
\begin{equation}
a_{1}(t)=\lambda_{1}(t)e^{i\varphi_{1}(t)},\quad\quad b_{1}(t)=\lambda_{2}(t)e^{i\varphi_{2}(t)},
\end{equation}
where $\lambda_{i}(t)$ and $\varphi_{i}(t)$ are arbitrary real functions of time.

Since the Hamiltonians are time-dependent, in describing quantum dynamics we shall use the Schr\"odinger picture in which state vectors depend explicitly on time, but operators do not
\begin{equation}
i\hbar\frac{d}{dt}|\psi(t)\rangle=H(t)|\psi(t)\rangle.\label{Schr}
\end{equation}
Thus, in order to study the time evolution of the states of Hamiltonians $H(t)_{su(1,1)}$ and $H(t)_{su(2)}$, we will use the time-dependent nontrivial invariant Hermitian operator $I(t)$ \cite{Lewis1,Lewis2}, which satisfies the conditions
\begin{equation}
i\frac{\partial}{\partial t}I(t)+[I(t),H(t)]=0.\label{inavariante}
\end{equation}
By using the $SU(1,1)$ and $SU(2)$ time-dependent displacement operators $D(\xi_{i}(t))$ given by the expressions (\ref{do}), with $\xi_{i}(t)=-\frac{1}{2}\theta_{i}(t)e^{-i\gamma_{i}(t)}$ and where now  $\theta_{i}(t)$ and $\gamma_{i}(t)$ are arbitrary real functions of time, we can define the invariant operators $I(t)_{su(1,1)}$ and $I(t)_{su(2)}$ as (see reference \cite{Lai})
\begin{equation}
I(t)_{su(1,1)}=D(\xi_{1}(t))K_{0}D^{\dag}(\xi_{1}(t)),\quad\quad I(t)_{su(2)}=D(\xi_{2}(t))J_{0}D^{\dag}(\xi_{2}(t)),\label{opinva}
\end{equation}
or explicitly as
\begin{equation}
I(t)_{su(1,1)}=\cosh(\theta_{1})K_{0}+\frac{\sinh(\theta_{1})}{2}e^{-i\gamma_{1}}K_{+}+\frac{\sinh(\theta_{1})}{2}e^{i\gamma_{1}}K_{-},\label{I1}
\end{equation}
and
\begin{equation}
I(t)_{su(2)}=\cos(\theta_{2})J_{0}+\frac{\sin(\theta_{2})}{2}e^{-i\gamma_{2}}J_{+}+\frac{\sin(\theta_{2})}{2}e^{i\gamma_{2}}J_{-}.\label{I2}
\end{equation}
From the condition of equation (\ref{inavariante}) and the invariant operators $I(t)_{su(1,1)}$ and $I(t)_{su(2)}$, the time-dependent physical parameters $b_{0}(t)$, $a_{0}(t)$, $\lambda_{i}(t)$ and $\varphi_{i}(t)$ are related to parameters $\theta_{i}(t)$ and $\gamma_{i}(t)$ as follows
\begin{equation}
\dot{\theta_{1}}=-2\lambda_{1}\sin(\varphi_{1}+\gamma_{1}),\quad\quad(\dot{\gamma_{1}}-a_{0})\sinh(\theta_{1})=-2\lambda_{1}\cosh(\theta_{1})\cos(\varphi_{1}+\gamma_{1}),\label{cond1}
\end{equation}
\begin{equation}
\dot{\theta_{2}}=-2\lambda_{2}\sin(\varphi_{2}+\gamma_{2}),\quad\quad(\dot{\gamma_{2}}-b_{0})\sin(\theta_{2})=-2\lambda_{2}\cos(\theta_{2})\cos(\varphi_{2}+\gamma_{2}).\label{cond2}
\end{equation}

On the other hand, the transformations of the generators $\{K_{0},K_{\pm}\}$ and $\{J_{0},J_{\pm}\}$ under its respective time-dependent displacement operators $D(\xi(t))$ remain unchanged and are given by the expressions (\ref{transformations}). In addition, by using of the BCH identity (equation (\ref{BCH})) the operator $i\frac{\partial}{\partial t}$ is transformed under the $SU(1,1)$ time-dependent displacement operators $D(\xi_{1})$ as
\begin{equation}
D_{1}^{\dag}(t)\left(i\frac{\partial}{\partial t}\right)D_{1}(t)=i\frac{\partial}{\partial t}+\dot{\gamma_{1}}(\cosh(\theta_{1})-1)K_{0}-\frac{e^{-i\gamma_{1}}}{2}\left(\dot{\gamma_{1}}\sinh(\theta_{1})+i\dot{\theta_{1}}\right)K_{+}-\frac{e^{i\gamma_{1}}}{2}\left(\dot{\gamma_{1}}\sinh(\theta_{1})-i\dot{\theta_{1}}\right)K_{-},\label{dt1}
\end{equation}
Analogously, the transformation of $i\frac{\partial}{\partial t}$ under the $SU(2)$ time-dependent displacement operator $D(\xi_{2})$ is given by
\begin{equation}
D_{2}^{\dag}(t)\left(i\frac{\partial}{\partial t}\right)D_{2}(t)=i\frac{\partial}{\partial t}+\dot{\gamma_{2}}(\cos(\theta_{2})-1)J_{0}-\frac{e^{-i\gamma_{2}}}{2}\left(\dot{\gamma_{2}}\sin(\theta_{2})+i\dot{\theta_{2}}\right)J_{+}-\frac{e^{i\gamma_{2}}}{2}\left(\dot{\gamma_{2}}\sin(\theta_{2})-i\dot{\theta_{2}}\right)J_{-}.\label{dt2}
\end{equation}

As it is shown in reference \cite{Lewis2}, if the eigenstates of the invariant operator satisfy the Schr\"odinger equation its eigenvalues are real. Therefore, given that $K_{0}|k,n\rangle=(k+n)|k,n\rangle$ and $J_{0}|j,\mu\rangle=\mu|j,\mu\rangle$ we have
\begin{equation*}
D(\xi_{1})K_{0}|k,n\rangle=(k+n)D(\xi_{1})|k,n\rangle,\quad\quad\quad\quad D(\xi_{2})J_{0}|j,\mu\rangle=\mu D(\xi_{2})|j,\mu\rangle,
\end{equation*}
\begin{equation*}
 I(t)_{su(1,1)}D(\xi_{1})|k,n\rangle=(k+n)D(\xi_{1})|k,n\rangle,\quad\quad I(t)_{su(2)}D(\xi_{2})|j,\mu\rangle=\mu D(\xi_{2})|j,\mu\rangle.
\end{equation*}
Thus, the invariant operators $I(t)_{su(1,1)}$ and $I(t)_{su(2)}$ have as eigenstates, the states
$D(\xi_{1})|k,n\rangle=|\zeta_{1}(t),k,n\rangle$ and $D(\xi_{2})|j,\mu\rangle=|\zeta_{2}(t),j,\mu\rangle$ respectively, which are the $SU(1,1)$ and $SU(2)$ Perelomov number coherent states. These states are given by the expressions (\ref{PNCS}) and (\ref{PNCS2}) but now these are functions of time through the parameters
\begin{equation}
\zeta_{1}(t)=-\tanh\left(\frac{\theta_{1}(t)}{2}\right)e^{-i\gamma_{1}(t)},\quad\quad \eta_{1}=\ln(1-|\zeta_{1}(t)|^{2}),
\end{equation}
\begin{equation}
\zeta_{2}(t)=-\tan\left(\frac{\theta_{2}(t)}{2}\right)e^{-i\gamma_{2}(t)},\quad\quad \eta_{2}=\ln(1+|\zeta_{2}(t)|^{2}).
\end{equation}
Finally, the eigenvalues of $I(t)_{su(1,1)}$ are $(k+n)$ and the eigenvalues of $I(t)_{su(2)}$ are $\mu$.

Moreover, if the states $|\psi(t)\rangle_{su(1,1)}$ and $|\psi(t)\rangle_{su(2)}$ satisfy the relationship (\ref{Schr}) for the Hamiltonians $H(t)_{su(1,1)}$ and $H(t)_{su(2)}$, these states can be expanded through the sates $|\zeta_{1}(t),k,n\rangle$ and $|\zeta_{2}(t),j,\mu\rangle$  in the form
\begin{equation}
|\psi(t)\rangle_{su(1,1)}=\sum_{n}a_{n}e^{i\alpha^{(1)}_{n}}|\zeta_{1}(t),k,n\rangle,\quad\quad
|\psi(t)\rangle_{su(2)}=\sum_{j}a_{j}e^{i\alpha^{(2)}_{j}}|\zeta_{2}(t),j,\mu\rangle,
\end{equation}
where according to reference \cite{Lewis2} the phase $\alpha$ is given as
\begin{equation}
\alpha=\int_{0}^{t}dt'\langle \lambda,\kappa|i\hbar\frac{\partial}{\partial t'}-H(t')|\lambda,\kappa\rangle.\label{T-phase}
\end{equation}
Here, $|\lambda,\kappa\rangle$ are the eigenstates and $\lambda$ are the eigenvalues of the invariant operator $I(t)$.
Therefore, the phase of the eigenstate $|\zeta_{1}(t),k,n\rangle$ and $|\zeta_{2}(t),j,\mu\rangle$ in a non-adiabatic process is given by
\begin{equation}
\alpha^{(1)}_{n}=(n+k)\int_{0}^{t}\left[(\dot{\gamma_{1}}+a_{0})(\cosh(\theta_{1})-1)+2\lambda_{1}\cos(\gamma_{1}+\varphi_{1})\sin(\theta_{1})-a_{0}\right]dt',\label{nophase1}
\end{equation}
\begin{equation}
\alpha^{(2)}_{j}=\mu\int_{0}^{t}\left[(\dot{\gamma_{2}}-b_{0})(\cos(\theta_{2})-1)-2\lambda_{2}\cos(\gamma_{2}+\varphi_{2})\sin(\theta_{2})-a_{0}\right]dt'.\label{nophase2}
\end{equation}
Unlike in a non-adiabatic process, in an adiabatic process we have that $\dot{\theta_{i}}=\dot{\gamma_{i}}=0$ and so the expressions (\ref{cond1}) and (\ref{cond2}) become
\begin{equation}
\varphi_{1}+\gamma_{1}=n\pi,\quad\quad\tanh(\theta_{1})=\frac{2\lambda_{1}}{a_{0}}\cos(n\pi),\label{1cond1}
\end{equation}
\begin{equation}
\varphi_{2}+\gamma_{2}=n\pi,\quad\quad \tan(\theta_{2})=\frac{2\lambda_{2}}{b_{0}}\cos(n\pi).\label{2cond2}
\end{equation}
If we set $n=1$, the above conditions are reduced to the time-dependent versions of the expressions  (\ref{parame1}) and (\ref{parame2}). Therefore, in an adiabatic process the phase of the states $|\zeta_{1}(t),k,n\rangle$ and $|\zeta_{2}(t),j,\mu\rangle$ are reduced to
\begin{equation}
\alpha^{(1)}_{n}=-(n+k)\int_{0}^{t}\sqrt{a_{0}(t')^{2}-4\lambda_{1}^{2}(t')}dt',\label{dphase1}
\end{equation}
\begin{equation}
\alpha^{(2)}_{j}=-\mu\int_{0}^{t}\sqrt{a_{0}^{2}(t')+4\lambda_{2}^{2}(t')}dt'.\label{dphase2}
\end{equation}
These are known as the dynamical phases and are defined as
\begin{equation}
\dot{\epsilon}_{n}=\langle\lambda,\kappa|H(t')|\lambda,\kappa\rangle\label{dyphase},
\end{equation}
while the Berry phase is written as
\begin{equation}
\dot{\gamma}_{\kappa}=i\langle\lambda,\kappa|\frac{\partial}{\partial t}|\lambda,\kappa\rangle.\label{berphase}
\end{equation}
Thus, the Berry phases of the states $|\zeta_{1}(t),k,n\rangle$ and $|\zeta_{2}(t),j,\mu\rangle$ are obtained in the adiabatic limit as follows
\begin{equation}
\gamma^{(1)}_{n}(T)=(n+k)\oint (\cosh(\theta_{1})-1)d\varphi_{1},\label{Berrysu11}
\end{equation}
\begin{equation}
\gamma^{(2)}_{n}(T)=\mu\oint (\cos(\theta_{2})-1)d\varphi_{2},
\end{equation}
where $T$ denotes the period. It is obvious that in these cases the Berry phases do not depend on an explicit form of the functions $\varphi_{1}(t)$ or $\varphi_{2}(t)$.

\section{$SU(1,1)$ and $SU(2)$ Berry phase of the Tavis-Cummings model}

In this Section, we shall consider a trilinear time-dependent Hamiltonian which vary slowly in the time. This Hamiltonian can be considered as the model of Tucker and Walls \cite{Tucker1,Tucker2} where only three modes interact with each other \cite{Mishkin,Agrawal,Abdalla}. This Hamiltonian has the form (with $\hbar=1$)
\begin{equation}
\hat{H}(t)=\omega_{1}(t)\hat{a}^{\dag}\hat{a}+\omega_{2}(t)\hat{b}^{\dag}\hat{b}+\omega_{3}\hat{c}^{\dag}\hat{c}+\lambda(t)(\hat{a}^{\dag}\hat{b}\hat{c}e^{-i\varphi(t)}+\hat{a}\hat{b}^{\dag}\hat{c}^{\dag}e^{i\varphi(t)}),\label{T-C}
\end{equation}
where the set of operators $\hat{a},\hat{a}^{\dag}$, $\hat{b},\hat{b}^{\dag}$ and $\hat{c},\hat{c}^{\dag}$ satisfy the bosonic algebra $[a,a^{\dag}]=[b,b^{\dag}]=[c,c^{\dag}]=1$. Moreover, $\omega_{j}(t)$ with $j=1,2,3$, $\lambda(t)$ and $\varphi(t)$ are physical time-dependent constants that vary slowly.

In what follows it will be convenient to use the bosonic $su(1,1)$ and $su(2)$ Lie algebras realizations
\begin{align}
&\hat{K}_0=\frac{1}{2}\left(\hat{b}^{\dag}\hat{b}+\hat{c}^{\dag}\hat{c}+1\right),\quad\hat{K}_+=\hat{b}^{\dag}\hat{c}^{\dag}\quad\hat{K}_-=\hat{b}\hat{c}\quad N_d=\hat{b}^{\dag}\hat{b}-\hat{c}^{\dag}\hat{c},\\&
\hat{J}_0=\frac{1}{2}\left(\hat{a}^{\dag}\hat{a}-\hat{b}^{\dag}\hat{b}\right),\quad\quad\quad \hat{J}_+=\hat{a}^{\dag}\hat{b},\quad \hat{J}_-=\hat{b}^{\dag}\hat{a}\quad\hat{N}_s=\hat{a}^{\dag}\hat{a}+\hat{b}^{\dag}\hat{b},
\end{align}
such that these operators satisfy the commutation relations (\ref{algebra1}) and (\ref{algebra2}). Thus, in terms of these operators the Hamiltonian (\ref{T-C}) can be rewritten in the following forms
\begin{align}
\hat{H}(t)_{su(1,1)}=\omega_{1}(t)\hat{a}^{\dag}\hat{a}+(\omega_{2}(t)+\omega_{3}(t))K_{0}+\lambda(t)(\hat{a}^{\dag}K_{-}e^{-i\varphi(t)}+\hat{a}K_{+}e^{i\varphi(t)})+\frac{\omega_{2}(t)-\omega_{3}(t)}{2}N_{d}-\frac{\omega_{2}(t)+\omega_{3}(t)}{2},\label{Hsu11}
\end{align}
\begin{equation}
\hat{H}(t)_{su(2)}=\omega_{3}(t)\hat{c}^{\dag}\hat{c}+(\omega_{2}(t)-\omega_{1}(t))\hat{J}_{0}+\lambda(t)(\hat{c}\hat{J}_{+}e^{-i\varphi(t)}+\hat{c}^{\dag}\hat{J}_{-}e^{i\varphi(t)})+\frac{\omega_{1}(t)+\omega_{2}(t)}{2}\hat{N}_{s}.\label{Hsu2}
\end{equation}
Therefore, the Hamiltonian (\ref{T-C}) possesses the $SU(1,1)$ and $SU(2)$ symmetries. By considering the $SU(1,1)$ Hamiltonian of (\ref{Hsu11}) and using the theory developed in Section 3, we have that the exact solution of the Schr\"odinger equation is obtained in terms of the eigenstates of the invariant operator $\hat{I}(t)_{su(1,1)}=D(\xi_{1})K_{0}D^{\dag}(\xi)$
\begin{equation}
|\psi(t)\rangle_{su(1,1)}=\sum_{n}a_{n}e^{i\alpha^{(1)}_{n}}|\zeta_{1}(t),k,n\rangle\otimes|n_{a}\rangle.
\end{equation}
In this expression, $|n_{a}\rangle$ are the states of the one-dimensional harmonic oscillator and the states $|\zeta_{1}(t),k,n\rangle=D(\xi_1)|k,n\rangle$ are given in terms of the quantum numbers $n_{l}$ and $m_{n}$ as
\begin{align}
\psi_{\zeta_{1},n_{l},m_n}=&\sqrt{\frac{2\Gamma(n_{l}+1)}{\Gamma(n_{l}+m_n+1)}}\frac{(-1)^{n_{l}}}{\sqrt{\pi}}e^{im_n\phi}\frac{(-\zeta_{1}^{*})^{n_{l}}(1-|\zeta_{1}|^{2})^{\frac{m_n}{2}+\frac{1}{2}}(1+\sigma)^{n_{l}}}{(1-\zeta_{1})^{m_n+1}}\nonumber\\&\times e^{-\frac{\rho^{2}(\zeta_{1}+1)}{2(1-\zeta_{1})}}\rho^{m_n}L_{n_{l}}^{m_m}\left(\frac{\rho^{2}\sigma}{(1-\zeta_{1})(1-\sigma)}\right),\label{eigen2}
\end{align}
where $\zeta_{1}(t)=-\tanh(\frac{\theta_{1}(t)}{2})e^{-i\gamma_{1}(t)}$ and $\sigma(t)$ is defined as
\begin{equation*}
\sigma(t)=\frac{1-|\zeta_{1}(t)|^{2}}{(1-\zeta_{1}(t))(-\zeta_{1}^{*}(t))}.
\end{equation*}

On the other hand, the total phase $\alpha^{(1)}_{n}$ is given by the expression (\ref{T-phase}), which consist of the usual dynamical phase $\epsilon^{(1)}_{n}$ and the Berry phase $\gamma^{(1)}(T)_{n}$. Thus, the dynamical phase of the $SU(1,1)$ Tavis-Cummings model is given by
\begin{align}
\epsilon^{(1)}_{n}(t)&=-\int_{0}^{t}\langle n_{a}|\otimes \langle\zeta_{1}(t),k,n|H_{su(1,1)}|\zeta_{1}(t),k,n\rangle\otimes|n_{a}\rangle dt'\\&=\int_{0}^{t}\langle n_{a}|\otimes \langle k,n|H'_{su(1,1)}|k,n\rangle\otimes|n_{a}\rangle dt',\label{su11dyph}
\end{align}
where $H'_{su(1,1)}=D^{\dag}H_{su(1,1)}D$. Taking into account the adiabatic limit and by using the relationship $\hat{a}^{\dag}\frac{1}{\hat{a}^{\dag}}=1-|0\rangle\langle 0|$, the equations (\ref{cond1}) become
\begin{equation}
\tanh(\theta_{1}(t))=\frac{2\lambda(t)\sqrt{\hat{a}^{\dag}\hat{a}}}{\omega_{2}(t)+\omega_{3}(t)},\quad\quad\gamma_{1}(t)=\frac{i}{2}\ln\left[\frac{\hat{a}}{\hat{a}^{\dag}}]\right]-\varphi(t),
\end{equation}
which are the expression of the coherent state parameters in the reference \cite{Nos3}. Now, if we impose that $n_{a}\gg 1$ and use the following relationship
\begin{equation}
\frac{1}{\hat{a}^{\dag}}|n_{a}\rangle=(1-\delta_{n_{a},0})\frac{1}{\sqrt{n_{a}}}|n_{a}+1\rangle,
\end{equation}
we have that
\begin{equation}
\langle\tanh(\theta_{1}(t))\rangle_{a}=\frac{2\lambda(t)\sqrt{n_{a}}}{\omega_{2}(t)+\omega_{3}(t)},\quad\quad\langle\gamma_{1}(t)\rangle_{a}=-\varphi(t).
\end{equation}
Therefore, the dynamical phase of the $SU(1,1)$ Tavis-Cummings model in the adiabatic limit is
\begin{equation}
\epsilon^{(1)}_{n}(t)=-\int_{0}^{t} E(t')_{su(1,1)}dt',\label{phase1}
\end{equation}
where explicitly the term $E(t)_{su(1,1)}$ is given by
\begin{equation}
E(t)_{su(1,1)}=\sqrt{(\omega_{2}(t)+\omega_{3}(t))^{2}-4\lambda(t)^{2}n_{a}}\left(n_l+\frac{m_n}{2}+\frac{1}{2}\right)+\omega_{1}(t)n_{a}+\frac{\omega_{3}(t)-\omega_{2}(t)}{2}m_{n}-\frac{\omega_{2}(t)+\omega_{3}(t)}{2}.\label{ESU(1,1)}
\end{equation}
In the same way, from expression (\ref{Berrysu11}) we obtain for the Berry phase of the $SU(1,1)$ Tavis-Cummings model in the adiabatic limit the following result
\begin{equation}
\gamma^{(1)}_{n}(T)=\left(n_l+\frac{m_n}{2}+\frac{1}{2}\right)\langle n_{a}|(\cosh(\theta_{1})-1)|n_{a}\rangle\oint d\varphi,
\end{equation}
\begin{equation}
\gamma^{(1)}_{n_{l},n_{a},m_{n}}(T)=\left(2n_l+m_n+1\right)\frac{(\omega_{2}(t)+\omega_{3}(t))-\sqrt{(\omega_{2}(t)+\omega_{3}(t))^{2}-4\lambda(t)^{2}n_{a}}}{\sqrt{(\omega_{2}(t)+\omega_{3}(t))^{2}-4\lambda(t)^{2}n_{a}}}\pi.
\end{equation}

Now, form the $SU(2)$ Tavis-Cummings model the invariant operator of the Hamiltonian $H(t)_{su(2)}$ is given by
\begin{equation}
\hat{I}(t)_{su(2)}=D(\xi_{2})J_{0}D^{\dag}(\xi_{2}.
\end{equation}
The general solution of the Schr\"odinger equation is given by the eigenstates of the invariant operator $\hat{I}_{su(2)}$
\begin{equation}
|\psi(t)\rangle_{su(2)}=\sum_{j}a_{j}e^{i\alpha^{(2)}_{j}}|\zeta_{2}(t),j,\mu\rangle\otimes|n_{c}\rangle,
\end{equation}
where $|n_{c}\rangle$ are the states of the one-dimensional harmonic oscillator for the oscillation mode $\hat{c}$ and $|\zeta_{2}(t),j,\mu\rangle$ are the $SU(2)$ Perelomov number coherent states for the two-dimensional harmonic oscillator. The states $|\zeta_{2}(t),j,\mu\rangle$ in the coordinate space are given in terms of the quantum numbers $n_{l}$ and $m_{n}$ by
\begin{align}
\psi_{\zeta_{2},n_{l},m_{n}}=&\frac{e^{-\frac{\rho^{2}}{2}}}{\sqrt{\pi}}\sum_{s=0}^{n_{l}+n}\frac{\zeta_{2}^{s}}{s!}\sum_{n=0}^{n_{l}+m_{n}}\frac{(-\zeta_{2}^{*})^{n}}{n!}e^{\frac{\eta}{2}(m_{n}-2n)}e^{i(m_{n}-2n+2s)\varphi}(-1)^{n_{l}+n-s}\nonumber\\&\times\frac{\Gamma(n_{l}+n+1)}{\Gamma(n_{l}+m_{n}-n+1)}\left[\frac{2\Gamma(n_{l}+m_{n}+1)}{\Gamma(n_{l}+1)}\right]^{1/2}\rho^{(m_{n}-2n+2s)}L_{n_{l}+n-s}^{(m_{n}-2n+2s)}(\rho^{2}),\label{ESU(2)}
\end{align}
where $\zeta_{2}(t)=-\tan(\frac{\theta_{2}(t)}{2})e^{-i\gamma_{2}(t)}$.

On the other hand, the total phase $\alpha^{(2)}_{j}$ is splitted into the dynamical phase $\epsilon^{(2)}_{j}$
and the Berry phase $\gamma^{(2)}_{j}(T)$. From the relationship $\hat{c}^{\dag}\frac{1}{\hat{c}^{\dag}}=1-|0\rangle\langle 0|$, we obtain that in the adiabatic limit the equations (\ref{cond2}) become
\begin{equation}
\tan(\theta_{2}(t))=\frac{2\lambda(t)\sqrt{\hat{c}^{\dag}\hat{c}}}{\omega_{2}(t)-\omega_{1}(t)},\quad\quad\gamma_{2}(t)=\frac{i}{2}\ln\left[\frac{\hat{c}}{\hat{c}^{\dag}}]\right]+\varphi.
\end{equation}
If we consider that $n_{a}\gg 1$ and use the relationship
\begin{equation}
\frac{1}{\hat{c}^{\dag}}|n_{c}\rangle=(1-\delta_{n_{c},0})\frac{1}{\sqrt{n_{}}}|n_{c}+1\rangle,
\end{equation}
we obtain that $\langle\gamma_{2}(t)\rangle_{c}=\varphi(t)$. Therefore, the dynamical phase for the $SU(2)$ Tavis-Cummings model is given by
\begin{equation}
\epsilon^{(2)}_{j}(t)=-\int_{0}^{t} E(t')_{su(2)}dt',\label{su2dyphase2}
\end{equation}
where explicitly the term $E(t)_{su(2)}$ is
\begin{equation}
E_{su(2)}=\frac{1}{2}\sqrt{(\omega_{2}(t)-\omega_{1}(t))^{2}+4\lambda^{2}(t)n_{c}}{m_{n}}+(\omega_{2}(t)+\omega_{1}(t))(n_l+\frac{m_n}{2})+\omega_{3}(t)n_{c}.
\end{equation}
Finally, the Berry phase $\gamma^{(2)}(T)_{j}$ of the $SU(2)$ Tavis-Cummings model in the adiabatic limit is given by
\begin{equation}
\gamma^{(2)}_{j}(T)=\left(\frac{m_n}{2}\right)\langle n_{c}|(\cos(\theta_{2})-1)|n_{c}\rangle\oint d\varphi,
\end{equation}
\begin{equation}
\gamma^{(2)}_{m_{n},n_{c}}(T)=\left(\pi m_{n}\right)\frac{(\omega_{2}(t)-\omega_{1}(t))-\sqrt{(\omega_{2}(t)-\omega_{1}(t))^{2}+4\lambda(t)^{2}n_{c}}}{\sqrt{(\omega_{2}(t)-\omega_{3}(t))^{2}+4\lambda(t)^{2}n_{c}}}.
\end{equation}

\section{Concluding remarks}

In this paper we first introduced a method to diagonalize Hamiltonians with an $SU(1,1)$ and $SU(2)$ linear algebraic structure. For this method, we applied the similarity transformations of the displacement operator to the generators of each group. With these transformations we were able to obtain in a general way the Berry phase of time-dependent Hamiltonians with an $SU(1,1)$ and $SU(2)$ linear structure. Then, we introduced a trilinear time-dependent form of the Tavis-Cummings model, which possesses the $SU(1,1)$ and $SU(2)$ symmetry. Therefore, all the previous results allowed us to compute the $SU(1,1)$ and $SU(2)$ Berry phases for the Tavis-Cummings model with three modes of oscillation.

It is important to note that the method developed in this work not only allowed us to calculate the Berry phase exactly, but also allowed us to obtain the dynamical phase of the Tavis-Cummings model for each symmetry. Moreover, the general method developed to obtain the Berry phase and the dynamical phase can be applied to more Hamiltonians who have these symmetries.

\section*{Acknowledgments}

This work was partially supported by SNI-M\'exico, COFAA-IPN, EDI-IPN, EDD-IPN, SIP-IPN project number $20195316$.


\begin{thebibliography} {99}

\bibitem{Jay} E.T. Jaynes, and F.W. Cummings, Proc. IEEE 51, 89 (1963).

\bibitem{Haroche} S. Haroche, and J.M. Raimond, Exploring the Quantum: Atoms, Cavities and Photons, Oxford
University Press, Oxford, 2007.

\bibitem{Narozhny} J.H. Eberly, N.B. Narozhny, and J.J. Sanchez-Mondragon, Phys. Rev. Lett. 44, 1323 (1980).

\bibitem{Kuklinski} J.R. Kukli\'nski, and J. Madajczyk, Phys. Rev. A 37, 3175 (1988).

\bibitem{Short} R. Short, and L. Mandel, Phys. Rev. Lett. 51, 384 (1983).

\bibitem{Diedrich} F. Diedrich, and H. Walther, Phys. Rev. Lett. 58, 203 (1987).

\bibitem{Milonni} P.W. Milonni, J.R. Ackerhalt, and H.W. Galbraith, Phys. Rev. Lett. 50, 966 (1983).

\bibitem{Slosser} J.J. Slosser, P. Meystre, and S.L. Braunstein, Phys. Rev. Lett. 63, 934 (1989).

\bibitem{Gea} J. Gea-Banacloche, Phys. Rev. Lett. 65, 3385 (1990).

\bibitem{Phoenix} S.J.D. Phoenix, and P.L. Knight, Phys. Rev. Lett. 66, 2833 (1991).

\bibitem{Goy} P. Goy, J.M. Raimond, M. Gross, and S. Haroche, Phys. Rev. Lett. 50, 1903 (1983).

\bibitem{Brune} M. Brune et. al., Phys. Rev. Lett. 76, 1800 (1996).

\bibitem{Guerlin} C. Guerlin et. al., Nature 448, 889 (2007).

\bibitem{Dicke} R.H. Dicke, Phys. Rev. 93, 99 (1954).

\bibitem{TC} M. Tavis, and F.W. Cummings, Phys. Rev. 170, 379 (1968).

\bibitem{Bashir} M.A. Bashir, and M.S. Abdalla, Phys. Lett. A 204, 21 (1995).

\bibitem{Bogoliubov} N.M. Bogoliubov, R.K. Bullough, and J. Timonen, J. Phys. A: Math. Gen. 29, 6305 (1996).

\bibitem{Rybin} A. Rybin, G. Kastelewiczz, J. Timonen, and N.M. Bogoliubov, J. Phys. A: Math. Gen. 31, 4705 (1998).

\bibitem{Vadeiko} I.P. Vadeiko, G.P. Miroshnichenko, A.V. Rybin, and J. Timonen, Phys. Rev. A 67, 053808 (2003).

\bibitem{Lamata} L. Lamata, J. Le\'on, T. Sch{\"a}tz, E. Solano, Phys. Rev. Lett. 98, 253005 (2007).

\bibitem{Gerritsma} R. Gerritsma et al., Nature (London) 463, 68 (2010).

\bibitem{Lamata2} L. Lamata et al., New J. Phys. 13, 095003 (2011).

\bibitem{Retzker} A. Retzker, E. Solano, and B. Reznik, Phys. Rev. A 75, 022312 (2007).

\bibitem{Kopylov} W. Kopylov et al., Phys. Rev. A 92, 063832 (2015).

\bibitem{Sun} C. Sun, and N. Sinitsyn, Phys. Rev. A 94, 033808 (2016).

\bibitem{Nos1} D. Ojeda-Guill\'en, R.D. Mota, V.D. Granados, J. Math. Phys. 57, 062104 (2016).

\bibitem{Nos2} E. Chore$\tilde{n}$o, D. Ojeda-Guill\'en, M. Salazar-Ram\'irez, and V.D. Granados, Ann. Phys. 387, 121 (2017).

\bibitem{Nos3} E. Chore$\tilde{n}$o, D. Ojeda-Guill\'en, and V.D. Granados, J. Math. Phys. 59, 073506 (2018).

\bibitem{Nos4} E. Chore$\tilde{n}$o, D. Ojeda-Guill\'en, and V.D. Granados, Eur. Phys. J. D 72, 142 (2018).

\bibitem{Berry} M.V. Berry, Proc. R. Soc. Lond. A 392, 45 (1984).

\bibitem{Gerry} C.C. Gerry, Phys. Rev. A 39, 3204 (1989).

\bibitem{Xie} B.-H. Xie, S. Jin, W.-X. Yan, S.-Q. Duan, and X.-G. Zhao, Eur. Phys. J. D 30, 411 (2004).

\bibitem{Bu} S.-P. Bu, G.-F. Zhang, J. Liu, and Z.-Y. Chen, Phys. Scr. 78, 065008 (2008).

\bibitem{Thilagam} A. Thilagam, J. Phys. A: Math. Theor. 43, 354004 (2010).

\bibitem{Koc1} R. Ko\c{c}, M. Koca, and  H. T\"ut\"unc\"uler, J. Phys. A: Math. Gen 35, 9425 (2002).

\bibitem{Koc2} R. Ko\c{c}, O. \"Ozer, H. T\"ut\"unc\"uler, and R.G. Yildirim, Eur. Phys. J. B 59, 375 (2007).

\bibitem{Koc3} R. Ko\c{c}, H. T\"ut\"unc\"uler, M. Koca, and E. Ol\u{g}ar, Ann. Phys. 319, 333 (2005).

\bibitem{Vourdas} A. Vourdas, Phys. Rev. A 41, 1653 (1990).

\bibitem{Nos5} D. Ojeda-Guill\'en, M. Salazar-Ram\'irez, R.D. Mota, and V.D. Granados, J. Nonlinear Math. Phys. 23, 607 (2016).

\bibitem{Lewis1} H.R. Lewis Jr, J. Math. Phys. 9, 1976 (1968).

\bibitem{Lewis2} H.R. Lewis Jr, and W.B. Riesenfeld, J. Math. Phys. 10, 1458 (1969).

\bibitem{Lai} Y.-Z. Lai, J.-Q. Liang, H.J.W. M\"uller-Kirsten, and J.-G. Zhou, Phys. Rev. A 53, 3691 (1996).

\bibitem{Tucker1} J. Tucker, and D.F. Walls, Phys. Rev. 178, 2036 (1969).

\bibitem{Tucker2} J. Tucker, and D.F. Walls, Ann. Phys. NY 52, 1 (1969).

\bibitem{Mishkin} E.A. Mishkin, and D.F. Walls, Phys. Rev. 185, 1618 (1969).

\bibitem{Agrawal} G.P. Agrawal, and C.L. Mehta, J. Phys. A: Math. Gen. 7, 607 (1974).

\bibitem{Abdalla} M.S. Abdalla, E.M. Khalil, A.S.-F. Obada, J. Pe{\v{r}}ina, and J. K{\v{r}}epelka, AIP Advances 7, 015013 (2017).

\end{thebibliography}
\end{document}